\documentclass[prl,12pt]{revtex4-1}

\usepackage[english]{babel}
\usepackage[intlimits]{amsmath}
\usepackage[dvips]{graphicx}
\usepackage{amsfonts}
\usepackage{amssymb}
\usepackage{graphicx}
\usepackage{bm}

\begin{document}

\title{How Interactions Control Molecular Transport in Channels}

\author{Anatoly B. Kolomeisky and Karthik Uppulury}
\affiliation{Department of Chemistry, Rice University, Houston, TX 77005, USA}

\begin{abstract}

The motion of molecules across channels is critically important for understanding mechanisms of cellular processes. Here we investigate the mechanism of interactions in the molecular transport by analyzing exactly solvable discrete stochastic models. It is shown that the strength and spatial distribution of molecule/channel interactions can strongly modify the particle current. Our analysis indicates that the most optimal transport is achieved when the binding sites are near the entrance or exit of the pore. In addition, the role of intermolecular interactions is studied, and it is argued that an increase in  flux can be observed for some optimal interaction strength. The mechanism of these phenomena is discussed.

\end{abstract}

\maketitle

The role of biological channels is to support cellular processes by regulating fluxes of molecules and ions \cite{lodish_book,hille_book}. It is known that membrane protein pores are fast, efficient, selective, and their functioning is robust with respect to strong non-equilibrium fluctuations in the cellular environment. These observations are especially surprising since in many cases the molecular translocation does not involve the use of metabolic energy or conformational changes \cite{wickner_review}. Understanding molecular mechanisms of transport is one of the most fundamental problems in biological systems.

When molecules enter into the channel their motion is slowed down mostly due to entropic barriers, and additional forces are needed to overcome these barriers. There are increasing experimental evidences that high efficiency, speed and selectivity of many biological and artificial channels is a result of complex processes that involve molecule/pore and intermolecular interactions \cite{wickner_review,exp1,movileanu}. Recent high-resolution experiments on polypeptide translocations through protein nanopores \cite{movileanu} started to probe the effect of molecule/pore interactions at the single-molecule level. It was found that changing the location of the binding site in the pore significantly modified the polypeptide flux across the channel.  However, a comprehensive description of the role of interactions in the transport through the pores is still not available due to the strong biochemical and biophysical complexity of the translocation machinery and due to the lack of structural information \cite{wickner_review,movileanu}.

To uncover mechanisms of molecular transport across the nanopores several theoretical methods have been proposed \cite{chou,berezhkovskii,kolomeisky,zilman}.  The continuum models of the channel transport view the translocation  as  one-dimensional motion in an effective potential created by interactions with nanopores and with other molecules \cite{berezhkovskii}. A different approach utilized  discrete-state stochastic models in which the translocation dynamics is analyzed as hopping between discrete binding sites in the channel \cite{kolomeisky,zilman}. Theoretical calculations show that both continuum and discrete approaches are closely related. Current theoretical models provide a reasonable description of some features of transport processes in the nanopores  \cite{berezhkovskii,kolomeisky,zilman}. In this Letter, we analyze theoretically the effect of interactions on the molecular transport across channels. Specifically, we address the question of how the translocation dynamics is modified by the strength and spatial distribution of the binding sites, and also by the intermolecular interactions.

We consider transport of molecules in the nanopore as an effective one-dimensional motion along the discrete lattice of binding sites as illustrated in Fig. 1. There are $N$ binding sites in the channel, and concentrations of molecules to the left or right of the channel are equal to $c_{1}$ and $c_{2}$, respectively.  The molecule can move into the channel from the left (right) with the rate $u_{0}=k_{on}c_{1}$ ($w_{0}=k_{on} c_{2}$); and the particle can move out of the channel with rates $w_{1}$ and $u_{N}$: see Fig. 1. In the nanopore the molecule at the site $j$ ($j=1,2,\cdots,N$) can jump forward (backward) with the rate $u_{j}$ ($w_{j}$). First, consider the situation when only one particle can be found in the channel. The probability to find the molecule at  site $j$ at time $t$ is given by a function $P_{j}(t)$, and the translocation dynamics is fully described by a set of master equations,   
\begin{equation}
\frac{dP_j(t)}{dt}= u_{j-1}P_{j-1}(t)+ w_{j+1}P_{j+1}(t)- (u_j+w_j)P_j(t)
\end{equation}
for $j=1,\cdots,N$; while $P_{0}(t) \equiv P_{N+1}(t)= 1- \sum_{j=1}^{N} P_{j}(t)$ describes the completely empty channel at the time $t$ \cite{kolomeisky}. We have shown earlier \cite{kolomeisky} that this model with $N$ binding sites can be solved exactly at $t \rightarrow \infty$ by mapping it into a single-particle random walk model on an infinite periodic lattice (with a period equal to $N+1$). Specifically, for the uniform channel with zero particle concentration to the right of the pore ($w_{0}=0$)  the expression for the particle current is given by
\begin{equation}\label{J0}
J_0(N)= \frac{u u_0}{(N+1)\left(u+\frac{N}{2}u_0\right)},
\end{equation}
where $u_{j}=w_{j}=u$ ($j=1,\cdots,N$). To quantify the effect of interactions we assume that in one of the binding sites, say $k$, the particle interacts with the pore with potential $\varepsilon$ that differs from other sites. The case of $\varepsilon >0$ corresponds to attractive interactions, while negative $\varepsilon$ describe the repulsive binding site. The transition rates near the special binding site must satisfy the detailed balance conditions which lead to
\begin{equation}
\frac{u'_{k-1}}{w'_{k}}=\frac{u_{k-1}}{w_{k}}x, \quad \frac{u'_k}{w'_{k+1}}=\frac{u_{k}}{w_{k+1}}\frac{1}{x},
\end{equation}
where $u_{k-1}$, $u_{k}$, $w_{k}$ and $w_{k+1}$ correspond to the uniform channel without special interactions, and we define $x= \exp(\varepsilon/k_{B}T)$. The corresponding explicit expressions  for transition rates can now be written as \cite{AR}, 
\begin{equation}
u'_{k-1}=u_{k-1}x^{\theta}, \quad  u'_k=u_{k}x^{\theta-1}, \quad w'_k=w_{k}x^{\theta-1}, \quad w'_{k+1}=w_{k+1}x^{\theta},
\end{equation}
where the coefficient $\theta$ ($0 \leq \theta \leq 1$) describes how the potential modifies the corresponding transition rates \cite{kolomeisky,AR}. Now flux in the channel with the binding site at position $k$ is equal to
\begin{equation}\label{J-single}
J_{k}(N)= \frac{uu_{0}}{u\left[2x^{-\theta}+N-1\right]+u_{0}\left[2(k-1)x^{-\theta}+x^{1-\theta}+(N-k)x+\frac{N(N-1)}{2}-k+1\right]}.
\end{equation}
The effect of interactions can be better understood by analyzing the ratio of particle currents,
\begin{equation} \label{ratio}
\frac{J_{k}(N)}{J_{0}(N)}=\frac{(N+1)\left[(u/u_{0})+\frac{N}{2})\right]}{(u/u_{0})\left[2x^{-\theta}+N-1 \right]+\left[2(k-1)x^{-\theta}+x^{1-\theta}+(N-k)x+\frac{N(N-1)}{2}-k+1\right]}.
\end{equation}
The curves presented in Fig. 2 show how  particle fluxes change depending on the position of the binding site. For attractive interactions the most optimal current is reached when the binding site is the last one ($k=N$), while it is better to have the repulsive site at the entrance ($k=1$) to accelerate the transport. It can be shown rigorously from Eq. (\ref{ratio}) that $\frac{\partial J_{k}(N)}{\partial k} > 0 $ for positive $\varepsilon$, and $J_{k}(N)$ is always a decreasing function  for negative $\varepsilon$. These observations can be understood in the following way. Putting the attractive binding site near the exit increases the probability of finding the particle here, which leads to higher chances to complete the translocation by exiting to the right. The repulsive site at the entrance serves as a barrier for the particles that already passed it, lowering the probability of unsuccessful excursions without the translocation. These results are in agreement with single-molecule experiments on translocation of polypeptides \cite{movileanu}. In these experiments the mutation in the biological nanopore that increased the molecule/pore interaction  have led to faster transport  when the mutation site was near the exit. It might also explain why so many biological channels have their binding sites at the entrance and at the exit positions since these distributions will optimize the overall fluxes \cite{chacinska02}. Our results can be easily extended to more complex potential with several attractive and repulsive sites, and it can be shown that the most optimal flux is reached when repulsive sites cluster together near the entrance, while attractive sites tend to stay closer to the exit. 

The strength of interactions can also affect the flux through the nanopore as shown in Fig. 3, in agreement with previous theoretical findings for the channel-facilitated molecular transport \cite{berezhkovskii,kolomeisky}. For any set of parameters there is an optimal interaction strength $\varepsilon^{*}$ that can be obtained from $\frac{\partial J_{k}(N)}{\partial x}(\varepsilon^{*})=0$, yielding
\begin{equation} \label{eq_optimal}
2\theta \left[\frac{u}{u_{0}}+k-1 \right]=(1-\theta) x +(N-k) x^{1+\theta}.
\end{equation}
Specifically, for the most optimal site $k=N$ (for attractive interactions) we have the following expression for the most optimal interaction strength, 
\begin{equation}
\varepsilon^{*} = k_{B}T \ln\left[\frac{2\theta}{1-\theta}\left(\frac{u}{u_0}+N-1\right)\right].
\end{equation}
From Eq. (\ref{eq_optimal}) it can be shown that for $\theta=0$ we have $\varepsilon^{*}= -\infty$ for any position of the binding site, while for $\theta=1$  one can obtain
\begin{equation}
\varepsilon^{*} = \frac{1}{2}k_{B}T \ln \left[\frac{2\left(\frac{u}{u_0}+k-1\right)}{N-k}\right].
\end{equation}

During the translocation across the channels more than one molecule can be found inside the nanopores and interactions between them might become important for the transport \cite{zilman}. Previous theoretical treatments considered the effect of the particle crowding \cite{zilman}, but assuming only hard-core exclusion interactions and neglecting correlations. To investigate explicitly the effect of intermolecular interactions we consider a specific $N=2$ model without molecule/pore interactions, as specified above, but allowing  more than one particle to be found in the pore. There is an energy cost associated with finding two particles next to each other. The configuration with 2 particles has an energy $\varepsilon$, with  $\varepsilon>0$ ($\varepsilon<0$) describing attractive (repulsive) interactions. There are four possible configurations in the channel as plotted in Fig. 4.  We label them as ($i,j$) with $i,j=0$ ($i,j=1$) for the empty (occupied) site. It should be noted that the rate to enter the half-filled configuration $u_{1}$ and the exit rate from the fully occupied state $u_{2}$ are related via the detailed balance,
\begin{equation}
\frac{u_1}{u_2}= \frac{u_0}{u} x,
\end{equation}
with $ x=\exp(\varepsilon/k_BT)$. The case $\varepsilon=0$ corresponds to the situation analyzed in Refs. \cite{zilman}.  This allows us to write explicit expressions,
\begin{equation}
u_1=u_0x^{\theta}, \quad  u_2=ux^{\theta-1},
\end{equation}
where the coefficient $0 \leq \theta \leq 1 $ again specifies how the inter-particle interaction modifies  these entrance and exit rates. We can define $P(i,j;t)$ as the probability to find channel in the state $(i,j)$ at time $t$, and temporal evolution of the system dynamics can be found by analyzing corresponding master equations. Solving these equations at large times yields the expression for the molecular flux,
\begin{equation}\label{J-multi}
J_{2}=\frac{uu_0(u+\frac{u_0}{2}x^{\theta})}{3u^2+\frac{u^2_0}{2}(x+x^{\theta})+uu_0(3+\frac{x^{\theta}}{2})}.
\end{equation}
In the limit of $\varepsilon \rightarrow -\infty$ only single molecules can be found in the channel and Eq. (\ref{J-multi}), as expected, reduces to Eq. (\ref{J0}) for $N=2$, $J_{1}=\frac{u u_{0}}{3(u+u_{0})}$. The effect of intermolecular interactions on the channel fluxes, as shown in Fig. 5, is rather complex. For $\theta=0$ the flux is always a decreasing function of the interaction, and the single-particle transport is the most optimal. For $\theta=1$ the trend is reversed: the stronger the interaction, the larger the molecular flux. However, for intermediate values of $0<\theta<1$ a non-monotonous dependence is observed with the flux reaching a maximum at some optimal interaction strength. The optimal interaction could be positive or negative depending on the parameters of the system. For attractive interactions the presence of the particle in the channel stimulates the entrance of another particle into the pore, but it slows down exiting of both particles from the channel. For repulsive interactions partially-filled channels serve as a barrier for the particle to enter, but simultaneously the entering particle accelerates the exiting of the particle inside. The combination of these processes explains the complex behavior in the channel with multiple particles. 

To summarize, we have investigated the effect of interactions on the molecular transport across channels. Using exactly-solvable discrete stochastic models, we have shown that the strength of the interaction, as well as the spatial distribution, are important parameters that can effectively control molecular translocations through nanopores. It was calculated that the largest particle current can be achieved when attractive sites are near the exit and/or repulsive sites are near the entrance. We have argued that the mechanism of how the interaction affect the transport across the channel is based on controling local concentration of particles in the channel. Attractive sites increase the probability to find the particles at these binding sites, while the repulsive sites work as barriers preventing particles already in the channel from moving back. Our theoretical picture agrees well with  single-molecule experiments on translocation of polypeptides \cite{movileanu}, and it might also explain distribution of binding sites in real biological channels \cite{chacinska02}. In addition, we have studied the role of intermolecular interactions in the transport through nanopores. It was found that at some interaction strength the particle flux can be increased to reach the maximum level. The complex behavior could be explained by the fact that particles already in the channel might catalyze or inhibit the entrance into the channel of other particles. The presented theoretical model presents a theoretical framework for investigating complex transport phenomena in biological and artificial channels, and it might serve as a first step for further studies that must include more realistic structural and biochemical information. 

We would like to acknowledge  support from the Welch Foundation (grant C-1559),  and the U.S. National Science Foundation (grant  ECCS-0708765).

\newpage

\noindent {\bf Figure Captions:} \\\\

\noindent Fig. 1. A schematic picture of the discrete stochastic model with $N$ binding sites for the translocation through the pore.  

\vspace{5mm}

\noindent Fig. 2. (color online) The ratio of particle currents for different positions of the special binding site for the channel with $N=10$ binding sites. Circles are for $\varepsilon/k_{B}T =5$, $u/u_{0}=0.1$ and $\theta=0.5$. Squares are for $\varepsilon/k_{B}T =-5$, $u/u_{0}=0.1$ and $\theta=0.5$. Triangles are for $\varepsilon/k_{B}T =5$, $u/u_{0}=10$ and $\theta=0.5$. Diamonds are for $\varepsilon/k_{B}T =5$, $u/u_{0}=0.1$ and $\theta=0.0$.

\vspace{5mm}

\noindent Fig. 3. (color online) The ratio of particle currents as a function of the interaction strength for the channel with $N=10$ binding sites. For the solid curve the parameters are  $u/u_{0}=0.1$, $k=1$ and $\theta=0.5$. For the dotted curve the parameters are  $u/u_{0}=0.1$, $k=10$ and $\theta=0.5$. For the dashed curve the parameters are  $u/u_{0}=0.1$, $k=10$ and $\theta=0.8$.  For the dash-dotted curve the parameters are  $u/u_{0}=0.1$, $k=5$ and $\theta=0.9$. 

\vspace{5mm}

\noindent Fig. 4. A general schematic picture for the channel with $N=2$ binding sites and with intermolecular interactions. Open circles describe empty sites, while filled circles denote the occupied sites. 

\vspace{5mm}

\noindent Fig. 5. (color online) Ratio of the particle currents as a function of the intermolecular interaction for the channel with $N=2$ binding sites. For the solid curve the parameters are  $u/u_{0}=0.1$ and $\theta=0$. For the dotted curve the parameters are  $u/u_{0}=0.1$  and $\theta=0.5$. For the dashed curve the parameters are  $u/u_{0}=0.1$ and $\theta=1$.  For the dash-dotted curve the parameters are  $u/u_{0}=10$ and $\theta=0.5$.

\newpage

\noindent \\\\\\

\begin{figure}[ht]
\unitlength 1in
\resizebox{3.375in}{3.5in}{\includegraphics{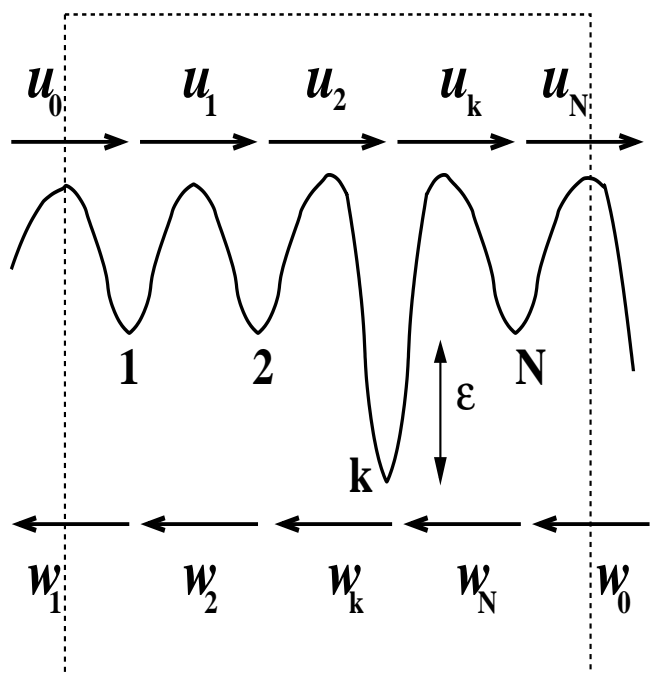}}
\vskip 0.3in
\caption{{\bf A. Kolomeisky and K. Uppulury, Physical Review Letters.}}
\end{figure}

\newpage

\begin{figure}[ht]
\unitlength 1in
\resizebox{3.375in}{2.5in}{\includegraphics{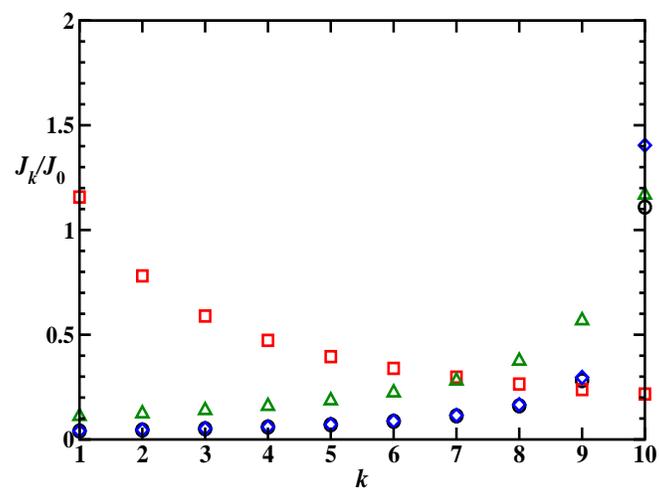}}
\vskip 0.3in
\caption{{\bf A. Kolomeisky and K. Uppulury, Physical Review Letters.}}
\end{figure}

\newpage

\begin{figure}[ht]
\unitlength 1in
\resizebox{3.375in}{2.5in}{\includegraphics{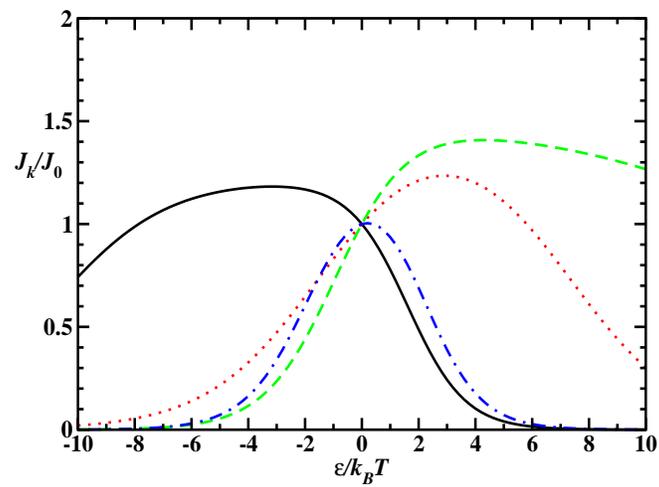}}
\vskip 0.3in
\caption{{\bf A. Kolomeisky and K. Uppulury, Physical Review Letters.}}
\end{figure}

\newpage

\begin{figure}[ht]
\unitlength 1in
\resizebox{3.375in}{3.2in}{\includegraphics{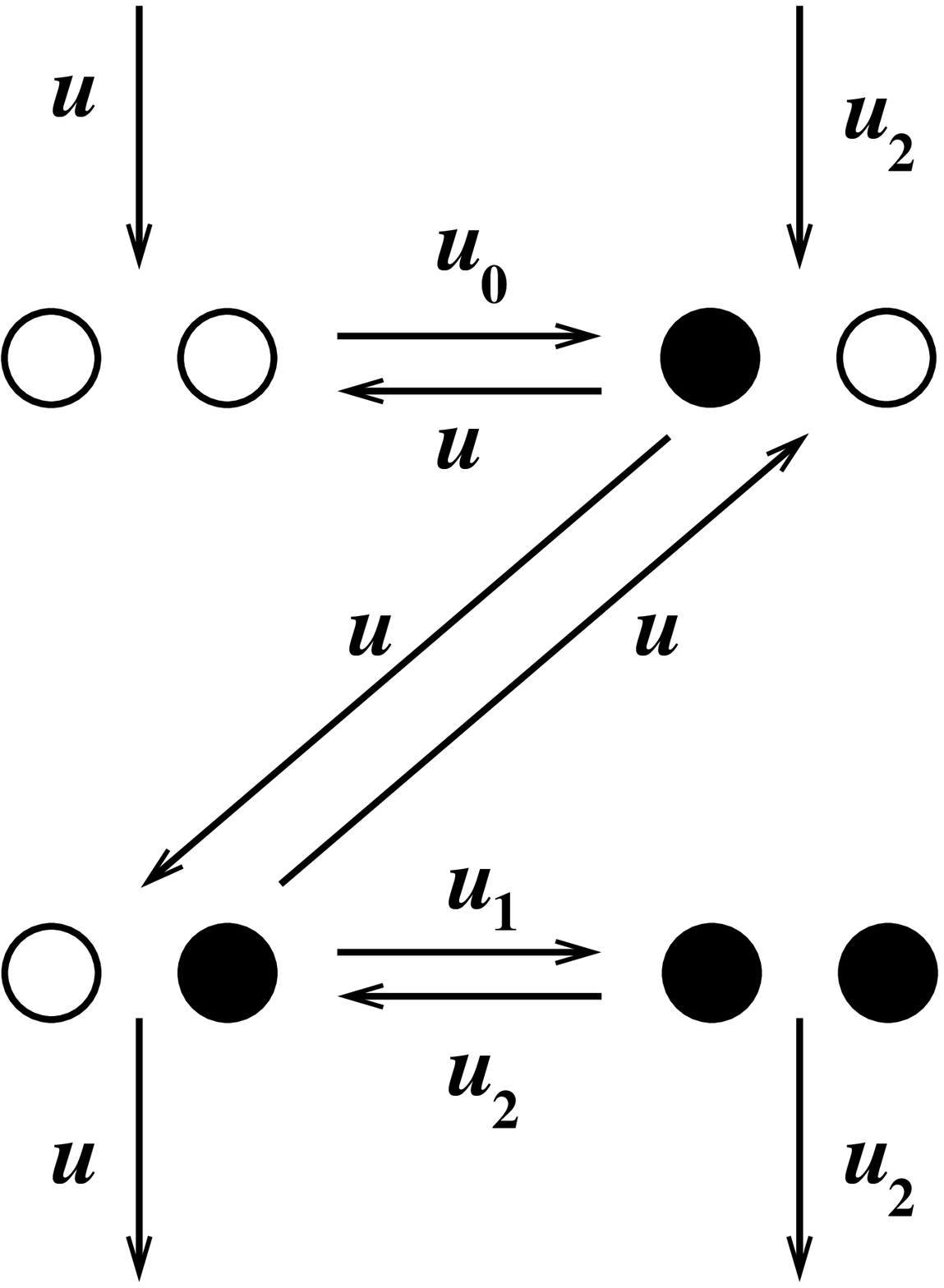}}
\vskip 0.3in
\caption{{\bf A. Kolomeisky and K. Uppulury, Physical Review Letters.}}
\end{figure}

\newpage

\begin{figure}[ht]
\unitlength 1in
\resizebox{3.375in}{2.5in}{\includegraphics{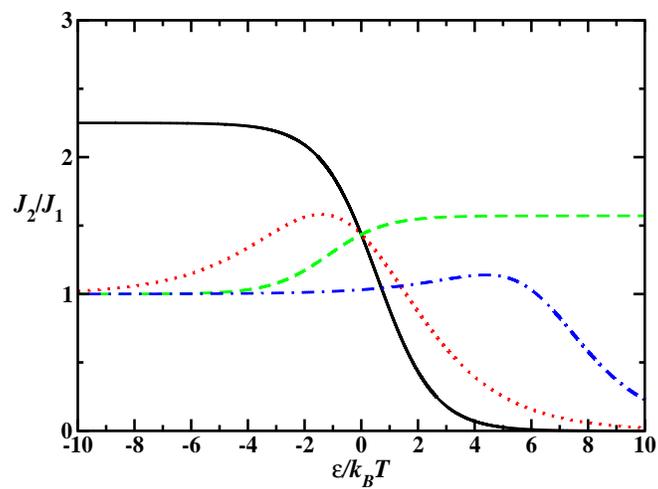}}
\vskip 0.3in
\caption{{\bf A. Kolomeisky and K. Uppulury, Physical Review Letters.}}
\end{figure}

\end{document}